\documentstyle[12pt]{article}
\begin{document}
\begin{flushright} 
OCU-PHYS 185 
\end{flushright} 
\begin{center}
{\Large {\bf Comment on fermion propagator in real-time 
quantum-field theory at finite temperature and density}} 
\end{center}
\hspace*{3ex}

\hspace*{3ex}

\hspace*{3ex}
\begin{center}
{\large {\sc A. Ni\'{e}gawa}\footnote{E-mail: 
niegawa@sci.osaka-cu.ac.jp}}
\end{center}

\begin{center}
{\normalsize\em Department of Physics, Osaka City University } \\ 
{\normalsize\em Sumiyoshi-ku, Osaka 558-8585, JAPAN} \\ 
\end{center}
\hspace*{3ex}

\hspace*{3ex} 

\hspace*{3ex}
\begin{center}
{\large {\bf Abstract}} \\ 
\end{center} 

Two forms are available for the fermion propagator at finite 
temperature and density. It is shown that, when one deals with a 
diquark-condensation-operator inserted Green function in hot and 
dense QCD, the standard form of the quark propagator does not work. 
On the other hand, another form of the quark propagator does work. 
\newpage
\subsection*{Introduction} 
A real-time thermal field theory (quantum-field theory at finite 
temperature and/or density) is formulated by introducing an oriented 
contour $C$ in a complex-time plane \cite{lan-le}, which starts from 
an initial time $\tau_0$ and ends at a final time $\tau_0 - i 
\beta$. Here $\beta = 1 / T$ and $\tau_0$ is an arbitrarily fixed 
point in the plane. Let $x^0$ and $y^0$ be any two points on $C$. If 
$x^0$ is nearer to $\tau_0$ on $C$ ($x^0$ is said to be \lq\lq past 
time'' when compared to $y^0$), $Im (x^0 - y^0) > 0$, $\theta_c (y^0 
- x^0) = 1$ and $\theta_c (x^0 - y^0) = 0$. A convenient contour 
usually employed in the literature consists of three straight 
segments: $C (\equiv C_R) = C_1 \oplus C_2 \oplus C_3$ where $C_1 = 
(\tau_0 = - \infty \to + \infty - i 0^+)$, $C_2 = (+ \infty - i 0^+ 
\to - \infty - 2 i 0^+)$, and $C_3 = (- \infty - 2 i 0^+ \to - 
\infty - i \beta)$. Hereafter, we refer this contour to as the 
real-time contour, since $C_R$ contains a real axis.  

We deal with a Dirac fermion field. The free Hamiltonian and the 
fermionic charge read, in respective order, 
\begin{eqnarray} 
H_0 & = &\int d^{\, 3} x \, \bar{\psi} (x) \left( - i 
\vec{\gamma} \cdot \vec{\nabla} + m \right) \psi(x) , 
\label{h0} \\ 
Q & = & \int d^{\, 3} x \, \psi^\dagger (x) \psi (x) . 
\end{eqnarray} 
Here $x^0 \in C$ and $\partial / \partial x^0$ is the 
$x^0$-derivative along $C$. The free propagator is defined as 
\begin{eqnarray} 
S_c (x - y) & = & Z^{- 1} \mbox{Tr} \left[ T_c \left( \psi (x) 
\bar{\psi} (y) \right) e^{- \beta (H_0 - \mu Q)} \right] , 
\label{shupp} \\ 
T_c \left( \psi (x) \bar{\psi} (y) \right) & = & \theta_c (x^0 - 
y^0) \psi (x) \bar{\psi} (y) - \theta_c (y^0 - x^0) \bar{\psi} (y) 
\psi (x) , 
\end{eqnarray} 
where $Z = \mbox{Tr} e^{- \beta (H_0 - \mu Q)}$. After Fourier 
transformation with respect to the spatial coordinates, we write 
\[ 
S_c (x^0 - y^0, \vec{p}) = \theta (x^0 - y^0) \, S_c^> (x^0 - y^0, 
\vec{p}) + \theta (y^0 - x^0) \, S_c^< (x^0 - y^0, \vec{p}) . 
\] 
\subsubsection*{Form 1 (F1) for the fermion propagator 
\cite{lan-le,le-b,land}:} 
The form to be given here is a standard one. We attach the 
superscript \lq\lq (1)'' to the F1 form here. The equation for the 
propagator reads (cf. Eqs.~(\ref{h0}) and (\ref{shupp})) 
\begin{equation} 
\left( i \gamma^0 \frac{\partial}{\partial x^0} - \vec{\gamma} 
\cdot \vec{p} - m \right) S^{(1)}_c (x^0 - y^0, \vec{p}) = \delta_c 
(x^0 - y^0) , 
\label{eq} 
\end{equation} 
where $\delta_c (x^0 - y^0) = \partial \theta_c (x^0 - y^0) / 
\partial x^0$. The operator $e^{\beta \mu Q}$ in Eq.~(\ref{shupp}) 
acts as distorting the boundary condition: 
\begin{equation} 
S^{(1)>}_c (x^0 - i \beta - y^0, \vec{p}) = - e^{- \beta \mu} 
S^{(1)<}_c (x^0 - y^0, \vec{p}) . 
\label{bc1} 
\end{equation} 
Solving Eq.~(\ref{eq}) under the condition (\ref{bc1}) yields 
\begin{eqnarray} 
S_c^{{(1)}>} (x^0 - y^0, \vec{p}) & = & \frac{- i}{2 \omega_p} 
\sum_{\tau = \pm} \left( \tau \gamma^0 \omega_p - \vec{\gamma} \cdot 
\vec{p} + m \right) [\theta (\tau) - n_\tau] e^{- i \tau \omega_p 
(x^0 - y^0)} , \nonumber \\ 
S_c^{{(1)}<} (x^0 - y^0, \vec{p}) & = & \frac{- i}{2 \omega_p} 
\sum_{\tau = \pm} \left( \tau \gamma^0 \omega_p - \vec{\gamma} \cdot 
\vec{p} + m \right) [\theta (- \tau) - n_\tau] e^{- i \tau \omega_p 
(x^0 - y^0)} , \nonumber \\ 
n_\pm & = & \frac{1}{e^{\beta (\omega_p \mp \mu)} + 1} , \;\;\;\;\;\;
\;\; \omega_p = \sqrt{p^2 + m^2} . 
\label{np} 
\end{eqnarray} 

Specification of so far arbitrary $C$ to the real-time contour $C_R$ 
leads to a standard two-component formalism of real-time thermal 
field theory (RT-TFT) \cite{lan-le,le-b,land,nie}. After Fourier 
transformation with respect to $x^0 - y^0$, we obtain for the matrix 
propagator $S_{i j}^{(1)}$: 
\begin{eqnarray} 
S_{i j}^{(1)} (P) & = & \left( 
{P\kern-0.1em\raise0.3ex\llap{/}\kern0.15em\relax} + m \right) 
\tilde{S}_{i j}^{(1)} (P) \;\;\;\;\;\;\;\; (i, j = 1, 2) c1 
\label{c1} 
\\ 
\tilde{S}_{11}^{(1)} (P) & = & \frac{1}{P^2 - m^2 + i 0^+} 
+ 2 \pi i N (p^0) \delta (P^2 - m^2) \\ 
& = & - \left[ \tilde{S}_{22}^{(1)} (P) \right]^* , \\ 
\tilde{S}_{12 (21)}^{(1)} (P) & = & - 2 \pi i \left[ \theta (\mp 
p^0) - N (p^0) \right] \, \delta (P^2 - m^2) , 
\label{10d} 
\\ 
N (p^0) & = & \theta (p^0) N_+ (p^0) + \theta (- p^0) N_- 
(p^0) , \;\;\;\;\;\; N_\pm (p^0) = \frac{1}{e^{\pm \beta (p^0 - 
\mu)} + 1} , 
\label{prop1} 
\end{eqnarray} 
where $P^\mu = (p^0, {\bf p})$. Here we have followed the so-called 
$|p^0|$-prescription \cite{nie}; $n_\pm \to N_\pm (p^0)$. 
\subsubsection*{Form 2 (F2) for the fermion propagator:} 
Here $H_0 - \mu Q$ is regarded as a free Hamiltonian. In place of 
Eqs.~(\ref{eq}) and (\ref{bc1}), we have, in respective order, 
\begin{eqnarray} 
&& \left( i \gamma^0 \frac{\partial}{\partial x^0} + \gamma^0 \mu - 
\vec{\gamma} \cdot \vec{\nabla} - m \right) S^{(2)}_c (x^0 - y^0, 
\vec{p}) = \delta_c (x^0 - y^0) , 
\label{eq2} \\ 
&& S^{(2)>}_c (x^0 - i \beta - y^0, \vec{p}) = - S^{(2)<}_c (x^0 - 
y^0, \vec{p}) . 
\label{bc2} 
\end{eqnarray} 
Comparing Eqs.~(\ref{eq2}) and (\ref{bc2}) with Eqs.~(\ref{eq}) and 
(\ref{bc1}), we see that 
\[ 
S^{(2)} (x^0 - y^0, \vec{p}) = e^{i \mu (x^0 - y^0)} S^{(1)} (x^0 - 
y^0, \vec{p}) . 
\] 
In \cite{le-b}, $S^{(2)}$ is denoted by $S'$. The RT-TFT matrix 
propagator $S_{i j}^{(2)}$ reads 
\begin{eqnarray} 
S_{i j}^{(2)} (p^0, \vec{p}) & = & S_{i j}^{(1)} (p^0 + \mu, p) 
\;\;\;\;\;\;\; (i, j = 1, 2) , 
\label{prop2} 
\end{eqnarray} 
where $S_{i j}^{(1)}$ is as in Eqs.~(\ref{c1})-(\ref{prop1}). 
\subsection*{Standard perturbation theory 
\cite{lan-le,le-b,land}} 
As far as standard perturbative calculations in RT-TFT are 
concerned, uses of the F1 form and of the F2 form yield the same 
result: 
\begin{itemize} 
\item 
A fermion loop (with loop momentum $P^\mu$): As can be seen from 
Eq.~(\ref{prop2}), changing the integration variable $p^0$ in the 
calculation using the C2 form as $p^0 \to p^0 - \mu$, we obtain the 
corresponding formula with the C1 form. 
\item 
An open fermion line with on-shell end points: Let $P$ be a 
momentum entering from the incoming end point into the diagram 
under consideration and $Q$ be a momentum going out from the 
outgoing end point. 
\begin{itemize} 
\item Computation with the C1 form: The wave function $F (\vec{p})$ 
[$G (\vec{p})$] is attached to the incoming [outgoing] line. $p^0$ 
[$q^0$] is the on-shell energy: $|p^0| = \omega_p = \sqrt{p^2 + 
m^2}$ [$|q^0| = \omega_q$]. For $p^0 > 0$ ($p^0 < 0$), $F (\vec{p}) 
= u (\vec{p})$ ($F (\vec{p}) = v (\vec{p})$), while, for $q^0 > 0$ 
($q^0 < 0$), $G (\vec{q}) = \bar{u} (\vec{q})$ ($\bar{v} (\vec{q})$). 
\item Computation with the C2 form: The wave function $F (\vec{p})$ 
[$G (\vec{p})$] is attached to the incoming [outgoing] line. $p^0 + 
\mu$ [$q^0 + \mu$] is the on-shell energy: $|p^0 + \mu| = \omega_p$ 
[$|q^0 + \mu| = \omega_q$]. For $p^0 + \mu> 0$ ($p^0 + \mu< 0$), $F 
(\vec{p}) = u (\vec{p})$ ($F (\vec{p}) = v (\vec{p})$), while, for 
$q^0 + \mu> 0$ ($q^0 + \mu < 0$), $G (\vec{q}) = \bar{u} (\vec{q})$ 
($\bar{v} (\vec{q})$). 
\end{itemize} 
\end{itemize} 
It can easily be seen that both schemes are equivalent also for an 
operator-inserted amplitude, as far as the operator consists of 
equal numbers of $\psi$ and $\bar{\psi}$ and of, if any, other 
fields. 
\subsubsection*{Mass insertion} 
It is instructive to see a mass-insertion \cite{nie}, i.e., a Green 
function in which inserted the operator ${\cal O}_M (z) = \bar{\psi} 
(z) \psi (z)$. We delete the superscripts \lq\lq $(i)$'' ($i = 1, 
2$). The ${\cal O}_M$-inserted Green function of our concern reads 
\[ 
\int_C d z^0 \int d^{\, 3} z \, \psi (x) {\cal O}_M (z) \bar{\psi} 
(y) = - \int_C d z^0 \int d^{\, 3} z \, S_c (x - z) \, S_c (z - y) . 
\] 

Fourier transforming with respect to spatial coordinates yields 
\begin{equation} 
- \int_{\tau_0}^{\tau_0 - i \beta} d z^0 \, S_c (x^0 - 
z^0, \vec{p}) \, S_c (z^0 - y^0, \vec{p}) \equiv F_c . 
\label{11} 
\end{equation} 
We see from Eqs.~(\ref{bc1}) and (\ref{bc2}) that 
\[ 
S^<_c (x^0 - \tau^0 + i \beta, \vec{p}) \, S^>_c (\tau^0 - i \beta- 
y^0, \vec{p}) = S^>_c (x^0 - \tau^0, \vec{p}) \, S^<_c (\tau^0 - 
y^0, \vec{p}) 
\] 
holds both for the F1 and F2. Then $F_c$, Eq.~(\ref{11}) is 
independent of $\tau_0$; $F_c = F_c (x^0 - y^0, \vec{p})$. 
Specifying $C$ to the real-time contour $C_R$, $F_c$ in 
Eq.~(\ref{11}) turns out to a $2 \times 2$ matrix quantity, whose 
Fourier transform (with respect to $x^0 - y^0$) is written as $F_{i 
j} (P)$ $(i, j = 1, 2)$. It is well known, or as can be directly 
shown using Eqs.~(\ref{c1})~-~(\ref{10d}) and (\ref{prop2}), that, 
for both C1 and C2, the same $F_{i j} (P)$ is also obtained through 
standard calculation using the RT-TFT Feynman rules: 
\[ 
F_{i j} (P) = - \sum_{k = 1}^2 (-)^{k - 1} \, S_{i k} (P) 
S_{k j} (P) \;\;\;\;\;\;\; (i, j = 1, 2) . 
\] 
From Eq.~(\ref{prop2}) follows 
\[ 
F^{(2)}_{i j} (p^0, \vec{p}) = F^{(1)}_{i j} (p^0 + \mu, 
\vec{p}) . 
\] 

Incidentally, one can derive the mass-derivative formula, as it 
should be, for both F1 and F2: 
\begin{eqnarray} 
\frac{\partial S_c (x^0 - y^0, \vec{p})}{\partial m} & = & - F_c 
(x^0 - y^0, \vec{p}) , 
\label{m1} 
\\ 
\frac{\partial S_{i j} (P)}{\partial m} & = & - F_{i j} (P) . 
\label{m2} 
\end{eqnarray} 
In deriving the formula (\ref{m2}), the so-called 
$|p^0|$-prescription \cite{nie} plays a role. Namely, if $N_\pm 
(p^0)$ in Eq.~(\ref{prop1}), which is in $S_{i j} (P)$, were $n_\pm$ 
(Eq.~(\ref{np})), Eq.~(\ref{m2}) would not hold. 
\subsection*{$\psi \psi$ and $\bar{\psi} \bar{\psi}$ insertions}
In recent years, possible diquark condensation in a quark matter has 
attracted much interest \cite{hara}. For dealing with this, \lq\lq 
condensate operators'' are introduced, such as, e.g., 
\[ 
{\cal O}_J (z) = J \sum_{a, b = 1}^2 \sum_{i, j = 1}^2 \epsilon_{a 
b 3} \epsilon_{i j} \psi^a_i (z) \gamma_5 C \psi^b_j (z) + 
\mbox{c.c.} . 
\] 
Here $J$ is a constant source, $a$ and $b$ are the color indices, 
$i$ and $j$ are the flavor indices [$i = 1$ stands for $u$ quark and 
$i = 2$ stands for $d$ quark], and $C$ here is the 
charge-conjugation matrix. The ${\cal O}_J$-inserted Green function 
reads 
\[ 
\int_C d z^0 \int d^{\, 3} z \, (\bar{\psi}^a_i (x))_\mu {\cal O}_J 
(z) (\bar{\psi}^b_j (y))_\nu = J \epsilon_{a b 3} \epsilon_{i j} 
\int_C d z^0 \int d^{\, 3} z \, \left[ S^T_c (z, x) \gamma_5 C S_c 
(z, y) \right]_{\mu \nu} , 
\] 
where and in the following, summations over the repeated indices are 
understood. The suffices $\mu$ and $\nu$ stand for the components of 
Dirac spinors, and $S^T_c$ is the transpose of $4 \times 4$ matrix 
function $S_c$. 

Fourier transformation with respect to the spatial coordinates 
yields 
\begin{equation} 
J \epsilon_{a b 3} \epsilon_{i j} \int_{\tau_0}^{\tau_0 - i \beta} d 
z^0 \, \left[ S_c (z^0 - x^0, \vec{p}) \right]_{\rho \mu} \left( 
\gamma_5 C \right)_{\rho \sigma} \left[ S_c (z^0 - y^0, - \vec{p}) 
\right]_{\sigma \nu} . 
\label{hosi} 
\end{equation} 
Now we observe that 
\begin{itemize} 
\item F1 form: 
\begin{eqnarray*} 
&& S_c^{(1)>} (\tau_0 - i \beta - x^0, \vec{p}) \, S_c^{(1)>} 
(\tau_0 - i \beta - y^0, - \vec{p}) \nonumber \\ 
&& \mbox{\hspace*{5ex}} = e^{- 2 \beta \mu} S_c^{(1)<} (\tau_0 - 
x^0, \vec{p}) \, S_c^{(1)<} (\tau_0 - y^0, - \vec{p}) . 
\label{aha} 
\end{eqnarray*} 
Then, Eq.~(\ref{hosi}) does depend on $\tau_0$. 
\item F2 form: 
\begin{eqnarray*} 
&& S_c^{(2)>} (\tau_0 - i \beta - x^0, \vec{p}) \, S_c^{(2)>} 
(\tau_0 - i \beta - y^0, - \vec{p}) \nonumber \\ 
&& \mbox{\hspace*{5ex}} = S_c^{(2)<} (\tau_0 - x^0, \vec{p}) \, 
S_c^{(2)<} (\tau_0 - y^0, - \vec{p}) . 
\label{aha1} 
\end{eqnarray*} 
Then, Eq.~(\ref{hosi}) is independent of $\tau_0$. 
\end{itemize} 
This statement also applies to $\int d z^0 \int d^{\, 3} z \, \psi 
(x) {\cal O}_J (z) \psi (x)$. Thus, in calculating the ($\psi 
\psi$)- and ($\bar{\psi} \bar{\psi}$)-inserted quantities, the F1 
form does not work and the F2 form should be employed. 

Straightforward computation with the F2 form shows that the 
following two computations yield the same result: 
\begin{itemize}  
\item 
First compute Eq.~(\ref{hosi}) for arbitrarily chosen contour $C$. 
Specify $C$ to the real-time contour $C_R$. Obtain a $2 \times 2$ 
matrix quantity through Fourier transformation 
with respect to 
$x^0 - y^0$. 
\item 
The corresponding $2 \times 2$ matrix quantity obtained by using the 
RT-TFT Feynman rules, the $(i j)$ element of which is 
\[ 
J \epsilon_{a b 3} \epsilon_{i j} (-)^{\gamma - 1} \left[ 
S^{(2)}_{\gamma \alpha} (P) \right]_{\rho \mu} \left( \gamma_5 C 
\right)_{\rho \sigma} \left[ S^{(2) \sigma \nu}_{\gamma \beta} (- 
P) \right]_{\sigma \nu} . 
\label{hosi1} 
\] 
\end{itemize}  
In a similar manner, the same statement holds for more general 
quantities: 
\begin{eqnarray*} 
&& \bar{\psi} (x) \prod_{i = 1}^{2 n} \left[ \int d^{\, 4} z_i \, 
{\cal O}_J (z_i) \right] \psi (y) , \nonumber \\ 
&& \bar{\psi} (x) \prod_{i = 1}^{2 n + 1} \left[ \int d^{\, 4} z_i 
\, {\cal O}_J (z_i) \right] \bar{\psi} (y) , \nonumber 
\end{eqnarray*} 
etc. It should be emphasized that the $|p^0|$-prescription plays an 
essential role here. 
\section*{Acknowledgments}
This work was supported in part by a Grant-in-Aide for Scientific 
Research ((C)(2) (No.~12640287)) of the Ministry of Education, 
Science, Sports and Culture of Japan. 

\end{document}